\documentstyle[twoside,fleqn,espcrc2,epsf,psfrag]{article}

\newcommand{\beqa}{\begin{eqnarray}}
\newcommand{\eeqa}{\end{eqnarray}}
\newcommand{\beq}{\begin{equation}}
\newcommand{\eeq}{\end{equation}}
\newcommand{\bc}{\begin{center}}
\newcommand{\ec}{\end{center}}
\newcommand{\lag}{2{\cal L(C)}}

\title{
A perturbative determination of O($a$) boundary
improvement coefficients for the Schr\"odinger Functional coupling
at 1-loop with improved gauge actions
\thanks{Talk presented by S.\ Takeda}}

\author{
Shinji Takeda,
Sinya Aoki and 
Kiyotomo Ide 
\\ 
Institute of Physics,
University of Tsukuba,
Tsukuba,
Ibaraki 305-8571,
Japan
}

\begin{document}
\pagestyle{empty}

\begin{abstract}
We perturbatively determine O($a$) boundary improvement
coefficients  at 1-loop 
for the Schr\"odinger Functional coupling
with improved gauge actions.
These coefficients are required to implement
the 1-loop O($a$) improvement in 
full QCD simulations for the coupling with the improved gauge actions.
To this order, lattice artifacts of the step scaling function (SSF) for 
each improved gauge action are also investigated.
Furthermore, we investigate 
the effect of the 1-loop O($a$) improvement to the SSF
in numerical simulations of the pure SU(3) gauge theory.
\end{abstract}

\maketitle

\section{Introduction}
\label{sec:introduction}

Recently CP-PACS/JLQCD Collaborations
have started the project for $N_f=3$ QCD simulations.
One of the targets in the project is to evaluate the strong coupling constant 
$\alpha_{\overline{\rm{MS}}}$ in $N_{\rm{f}}=3$ QCD  
using the the Schr\"odinger Functional (SF) scheme proposed by the ALPHA 
Collaboration \cite{sforiginal}.
In the project the RG improved gauge action\cite{Iwasaki}
has been employed to avoid the strong lattice artifacts found for
the plaquette gauge action for $N_f=3$ simulations\cite{okawa}.

In this report, as a first step toward the evaluation of 
$\alpha_{\overline{\rm{MS}}}$,
we study the SF coupling with improved gauge actions
including the plaquette and rectangle loops
in perturbation theory.
In particular,
the O($a$) boundary improvement coefficients 
are determined at 1-loop level for various improved gauge actions.
More detailed explanations concerning the 
perturbative calculations can be found
in \cite{takeda}.
Finally we examine the scaling behavior of the SSF
in numerical simulations of the pure SU(3) gauge theory
at the weak coupling region with tree or 1-loop $O(a)$
improvement coefficients.

\section{Set up}
\label{setup}

  \begin{figure}[b!]
 \vspace{-8mm}
   \begin{center}
       \psfragscanon
       \psfrag{tt}[][][1.5]{time}
       \psfrag{ss}[][][1.5]{space}
       \psfrag{T}[][][1.9]{$t=0$}
       \psfrag{ctp0}[][][1.9]{$c_0 c^P_{\rm{t}}(g_0^2)$}
       \psfrag{ctr1}[][][1.9]{$c_1$}
       \psfrag{ctr2}[][][1.9]{$c_1 c^R_{\rm{t}}(g_0^2)$}
       \scalebox{0.51}{\includegraphics{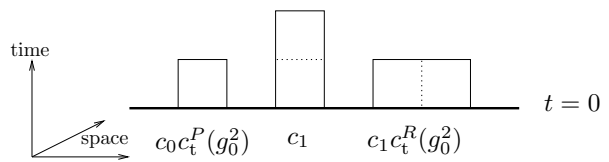}}
   \end{center}
 \vspace{-10mm}
\caption{The assignments near the boundary.
\label{boundary}}
  \end{figure}

We use the same SF set up 
as in the case of Wilson plaquette action\cite{su3},
except for the form of the action.
The action we employ is the improved gauge
actions including the plaquette(${\cal S}_0$) and rectangle(${\cal S}_1$)
 loops:
    \begin{equation}
      S_{\rm{imp}}[U] 
      =
      \frac{1}{g^2_0} 
      \sum_{i=0}^{1}
      \sum_{{\cal C} \in {\cal S}_i} W_i({\cal C},g^2_0) \lag.
     \label{eqn:impaction}
    \end{equation}
We refer to ref. \cite{takeda} for the unexplained notations.
The assignment of the weight factor $W_i({\cal C},g^2_0)$
is important to achieve the O($a$) improvement in the SF scheme.
We assign $W_i({\cal C},g^2_0)=c_i(c_0+8c_1$=$1)$
for all loops except for the ones near the boundary in the time direction.
The assignments at the $t=0$ boundary are shown in the Fig.\ref{boundary}.
At the $t=L$ boundary, similar assignments hold.
The O($a$) boundary improvement coefficients are given by
\beq
      c_0 c^P_{\rm{t}}(g^2_0)
      = 
      c_0 ( 1 + c^{P(1)}_{\rm{t}} g^2_0 + O(g^4_0) ), 
\eeq
\beq
      c_1 c^R_{\rm{t}}(g^2_0)
      = 
      c_1 ( 3/2 + c^{R(1)}_{\rm{t}} g^2_0 + O(g^4_0) ),
\eeq
where the leading terms necessary for the tree-level O($a$) improvement
is taken from \cite{aokiweisz}.
Our task is the determination of 1-loop terms
to achieve the 1-loop O($a$) improvement.

The SF coupling is defined through
the free energy of the system $\Gamma$
    \begin{equation}
      \bar{g}^2_{\rm{SF}}(L)
      =
      \left.
      (\Gamma^{\prime}_0 / \Gamma^{\prime})
      \right|_{\eta=\nu=0},
    \end{equation} 
where $\Gamma^{\prime}$ is the derivative with
respect to the background field parameter $\eta$.
$\Gamma^{\prime}_0$
is a normalization constant.
Using the redundant degree of freedom,
we take $c^{R(1)}_{\rm{t}}=2 c^{P(1)}_{\rm{t}}$
without loss of generality, 
then the perturbative expansion of
$\bar{g}^2_{\rm{SF}}(L)$
is given by
    \begin{equation}
      \bar{g}^2_{\rm{SF}}(L)
      =
      g^2_0 + m^{(1)}_1(L/a) g^4_0 + O(g^6_0), 
    \end{equation} 
    \begin{equation}
      m^{(1)}_1(L/a)
      =
    - (2a / L) c^{P(1)}_{\rm{t}}
    + m^{(0)}_1(L/a),
     \label{eqn:ctm1}
    \end{equation}
where $m^{(0)}_1(L/a)$
is the 1-loop correction
with the tree-level O($a$) boundary coefficients.
The detail of the calculation of $m^{(0)}_1(L/a)$
will be given in the next section.
The value of $c^{P(1)}_{\rm{t}}$
is determined by the improvement condition that
the dominant part of the scaling violation of
$m^{(1)}_1(L/a)$ should be proportional to $(a/L)^2$.

\section{Results}
\label{results}

In the following, we set $I=L/a$.
We evaluate the 1-loop correction $m^{(0)}_1(I)$
numerically 
for the Iwasaki action ($c_1 = -0.331$),
the L\"uscher-Weisz (LW) action 
($c_1 = -1/12$)
and the DBW2 action
($c_1 = -1.40686$)
in the range that $I = 6,\cdots,48$.

Symanzik's analysis
suggests that the $m^{(0)}_1(I)$ has
an asymptotic expansion such that
    \begin{equation}
      m^{(0)}_1(I)
      \stackrel{I \rightarrow \infty}{\sim}
      \sum^{\infty}_{n=0}
      [A_n + B_n \ln (I)]/(I)^n.
     \label{eqn:m1zenkin}
    \end{equation} 
Using the blocking method\cite{blocking},
we extracted the first few coefficients 
$A_0$, $B_0$, $A_1$, $B_1$ and estimated their errors.
As a check we numerically confirm the expected relations that $B_1 = 0$
and $B_0 = 2 b_0$, where $b_0$ is the 1-loop coefficient of the 
$\beta$ function, for various improved gauge actions.
Furthermore, as seen in the Table \ref{fig:A0A1},
$A_0$ extracted from (\ref{eqn:m1zenkin}) agree with $A_0^{\rm exp}$, 
estimated from the ratio of the $\Lambda$ parameter 
in \cite{su3,Iwasakiyoshie} and $A_0$ for the plaquette action\cite{su3}.
With these confidences in our computation, 
the main result, the 1-loop term of the O($a$) boundary improvement 
coefficient, is given by $c^{P(1)}_{\rm{t}}=A_1/2$,
where $A_1$ is found in Table \ref{fig:A0A1}.

\begin{table*} [] 
\vspace{-10mm}    
    \begin{center} 
     \begin{tabular}{cllll} 
       \hline \hline
         & plaquette action
         & Iwasaki action & LW action & DBW2 action 
       \\ \hline
           $A_0$            
         &\hspace{1.6mm} $0.36828215(13)$ 
         & $-0.2049015(4)$ 
         &\hspace{1.6mm} $0.136150567(6)$  
         & $-0.62776(8)$
       \\
           $A^{\rm{exp}}_0$ 
         & 
         & $-0.1999(24)$ 
         &\hspace{1.6mm} $0.13621(26)$ 
         & $-0.62$
       \\
           $A_1$
         & $-0.17800(10)$
         &\hspace{1.6mm} $0.30360(26)$
         & $-0.005940(2)$   
         &\hspace{1.6mm} $0.896(45)$
       \\
      \hline \hline
    \end{tabular}
    \end{center}
 \vspace{-5.5mm}    
   \caption{$A_0$, $A_1$
             for various gauge actions. 
             The values for plaquette action are taken
from \cite{su3}. Since the error of $A^{\rm{exp}}_0$ for 
DBW2 action is not given, the quoted
digits are of little significance.
    \label{fig:A0A1}}
\vspace{-1mm}    
    \end{table*}

\begin{figure}[t!]
\begin{tabular}{cc}
\hspace{-9mm}
\begin{minipage}{0.5\hsize}
\resizebox{55mm}{!}{\includegraphics{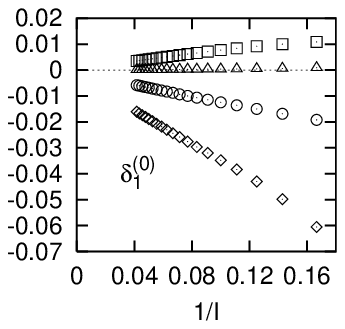}}
\end{minipage}
\hspace{2.0mm}
\begin{minipage}{0.5\hsize}
\resizebox{55mm}{!}{\includegraphics{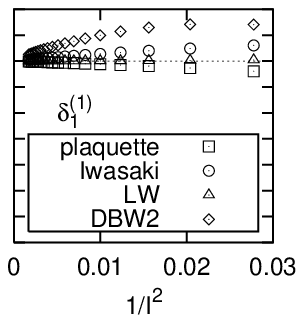}}
\end{minipage}
\end{tabular}
\vspace{-10mm}
\caption{The lattice cutoff dependence of the 1-loop deviations
for each degree of improvement and for various gauge actions.
} 
\label{fig:deviation}
\vspace{-7mm}
\end{figure}

Now let us discuss the lattice artifact of the
step scaling function (SSF).
The relative deviation of the lattice SSF $\Sigma(2,u,1/I)$
from the continuum SSF $\sigma(2,u)=g^2(2L)|_{g^2(L)=u}$ 
is given by
     \begin{eqnarray}
       \delta (2,u,1/I) 
       &=& 
         [\Sigma(2,u,1/I) - \sigma(2,u)]/ \sigma(2,u)
      \nonumber \\
      &=& 
       \delta^{(k)}_1(2,1/I) u  
     + O(u^2),
     \end{eqnarray} 
     \begin{equation}
       \delta^{(k)}_1(2,1/I)
       = 
       m_1^{(k)}(2I) 
     - m_1^{(k)}(I) 
     - 2 b_0 \ln(2) ,
      \label{eqn:deviation1}
     \end{equation}
where the $k$ denotes the degree of the improvement.
The results of $\delta^{(k)}_1(2,1/I)$,
including data of the plaquette action \cite{su3}
for comparison,
are given in Fig.\ref{fig:deviation}.
As is evident from Fig.\ref{fig:deviation},
the lattice artifact for the
RG improved gauge actions (Iwasaki or DBW2)
is comparable to or larger than that for the plaquette action,
while the LW action is the least affected by the lattice artifact.
It is also seen that the 1-loop deviation is much reduced by 
the 1-loop improvement, so that it is roughly proportional to $(a/L)^2$.

\vspace{0mm}

\section{Discussions}
\label{simulation}

\begin{figure}[t]
   \vspace{-2mm}
      \psfragscanon
      \psfrag{}[][][0.9]{}
      \scalebox{1.5}{\includegraphics{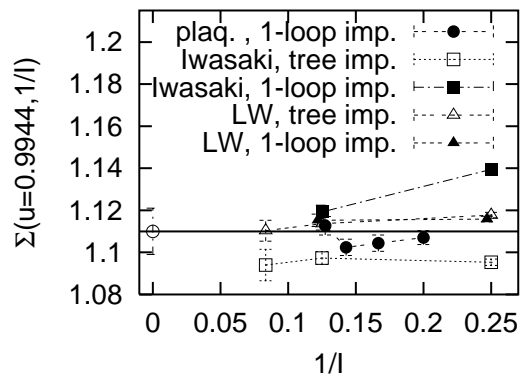}}
   \vspace{-14mm}
      \caption{The lattice cutoff dependence of SSF for various actions
at $u=0.9944$.
      \label{allaction}}
 \end{figure}

Using the O($a$) improvement coefficients obtained
in the previous section,
the CP-PACS collaboration is currently calculating the SSF non-perturbatively.
The preliminary results at the weak coupling region($u=0.9944$)
are given in the Fig.\ref{allaction} as a function of $a/L$, where
the data for the plaquette action\cite{su3}
are included for the comparison.
The scaling violation for the LW action 
is very small even in the tree improvement, and
the 1-loop improvement has almost no effect.
The scaling violation for the plaquette action with the 1-loop improvement
is also small.
On the other hand,
the Iwasaki action has larger scaling violation in the tree level improvement.
The 1-loop improvement reduces it at $I=8$, while
the scaling violation becomes larger at $I=4$.
This behavior may be understood in perturbation theory as follows.

By using the 1-loop deviation,
the lattice artifacts of SSF
can be estimated by $f_0$ and $f_1$ for each degree of improvement.
\beq
  f_0(1/I,u)
  =
  \sigma_{\rm cont}(2,u)
  (1+\delta^{(0)}_1(2,1/I) u),
\eeq
\beq
  f_1(1/I,u)
  =
  \sigma_{\rm cont}(2,u)
  (1+\delta^{(1)}_1(2,1/I) u),
\eeq
where $\sigma_{\rm cont}(2,u)=1.11$.
Fig.4 shows that not only the behavior but also the magnitude of
the scaling violation for the non-perturbative SSF
are consistent with the perturbative estimates.
In order to check whether this property holds at stronger coupling region,
we are currently calculating the SSF non-perturbatively at $u=2.4484$.

\

This work is supported in part by Grants-in-Aid for
Scientific Research from the Ministry of Education, 
Culture, Sports, Science and Technology.
(Nos. 13135204, 14046202, 15204015, 15540251).

 \begin{figure}[t]
   \vspace{-8mm}
      \psfragscanon
      \psfrag{}[][][0.9]{}
      \scalebox{1.5}{\includegraphics{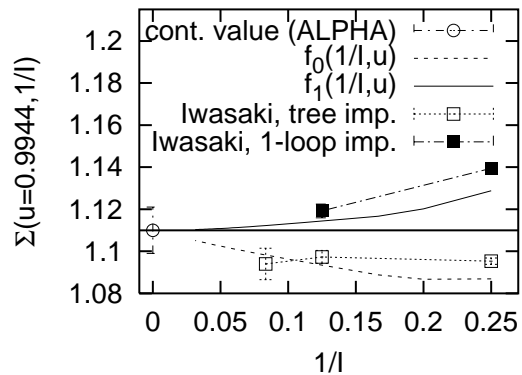}}
   \vspace{-14mm}
      \caption{The lattice cutoff dependence of SSF for the Iwasaki action
at $u=0.9944$.
      \label{Iwasakiaction}}
   \vspace{-3mm}
 \end{figure}

%

%
%


\begin{thebibliography}{l}



 \bibitem{sforiginal}
   M. L\"uscher, R. Narayanan, P. Weisz and U. Wolff, 
   Nucl. Phys. B384 (1992) 168
   
 \bibitem{Iwasaki}
   Y. Iwasaki,
   Nucl. Phys. B258 (1985) 141 ;
   Univ. of Tsukuba report UTHEP-118 (1983)

 \bibitem{okawa}
   S. Aoki et al. (JLQCD collaboration),
   Nucl. Phys. Proc. Suppl. 106 (2002) 263

 \bibitem{takeda}
   S. Takeda, S. Aoki and K. Ide,
   Phys. Rev. D68 (2003) 014505

 \bibitem{su3}
   M. L\"uscher, R. Sommer, P. Weisz and U. Wolff, 
   Nucl. Phys. B413 (1994) 481

 \bibitem{aokiweisz}
   S. Aoki, R. Frezzotti, and P. Weisz,
   Nucl.Phys. B540 (1999) 501





\bibitem{blocking}
  M. L\"uscher, P. Weisz, 
  Nucl. Phys. B266 (1986) 309

 \bibitem{Iwasakiyoshie}
   Y. Iwasaki, T. Yoshie, 
   Phys. Lett. 143B (1984) 449,
   Y. Iwasaki, S. Sakai, 
   Nucl. Phys. B248 (1984) 441,
   S. Sakai, T. Saito and A. Nakamura
   Nucl. Phys. B584 (2000) 528

\end{thebibliography}
\end{document}